\documentclass[
reprint,
superscriptaddress,
amsmath,
amssymb,
aps,
prl,
floatfix
]{revtex4-2}

\usepackage{graphicx}  
\usepackage{dcolumn}  
\usepackage{bm}   
\usepackage{comment}

\usepackage{hyperref}
\usepackage[dvipsnames]{xcolor}
\hypersetup{
    colorlinks=true,
    linkcolor=Blue,
    citecolor=Blue,   
   urlcolor=Blue
   }

\def\bea{\begin{eqnarray}}
\def\eea{\end{eqnarray}}
\def\beq{\begin{equation}}
\def\eeq{\end{equation}}

\def\la{\langle}
\def\ra{\rangle}

\def\nv{ {\hat{\mathbf{n}}}}
\def\rv{ {\mathbf{r}}}

\def\fv{ {\mathbf{f}}}
\def\Fv{ {\mathbf{F}}}
\def\vv{ {\mathbf{v}}}

\def\mupara{\mu_{\parallel}}

\def\b{\beta}

\begin{document}

\title{
Passivity-Driven Order--Disorder Transitions in Self-Aligning Active Matter
}

\author{Weizhen Tang}
\email[Weizhen Tang and Amir Shee contributed equally]{}
\affiliation{School of Systems Science, Beijing Normal University, Xinjiekouwaida Street 19, Beijing 100875, China}

\author{Amir Shee}
\email[Weizhen Tang and Amir Shee contributed equally]{}
\affiliation{Northwestern Institute on Complex Systems and ESAM, Northwestern University, Evanston, IL 60208, USA}
\affiliation{Department of Physics, University of Vermont, Burlington, Vermont 05405, USA}

\author{Zhangang Han}
\affiliation{School of Systems Science, Beijing Normal University, Xinjiekouwaida Street 19, Beijing 100875, China}

\author{Pawel Romanczuk}
\affiliation{Department of Biology, Humboldt Universität zu Berlin, Unter den Linden 6, Berlin 10099, Germany}
\affiliation{Research Cluster of Excellence ‘Science of Intelligence’, Berlin 10587, Germany}

\author{Yating Zheng}
\email[contact author:~]{yating.zheng@hu-berlin.de}
\affiliation{Department of Biology, Humboldt Universität zu Berlin, Unter den Linden 6, Berlin 10099, Germany}
\affiliation{Research Cluster of Excellence ‘Science of Intelligence’, Berlin 10587, Germany}

\author{Cristián Huepe}
\affiliation{School of Systems Science, Beijing Normal University, Xinjiekouwaida Street 19, Beijing 100875, China}
\affiliation{Northwestern Institute on Complex Systems and ESAM, Northwestern University, Evanston, IL 60208, USA}
\affiliation{CHuepe Labs, 2713 West Augusta Blvd \#1, Chicago, IL 60622, USA}

\date{\today}

\begin{abstract}
We study dense mixtures of passive and active self-aligning disks with isotropic or anisotropic mobility. We find that the passive fraction controls an order-disorder transition that is continuous in the isotropic case and discontinuous in the anisotropic one. A mean-field equation derived from the microscopic heading dynamics captures this dichotomy. Near the transition, both ordered regimes can exhibit multiple metastable oscillating or rotating states, depending on the spatial arrangement of passive particles and lattice defects, but with different transient dynamics: Systems with isotropic mobility visit multiple long-lived attractors during each simulation while systems with anisotropic mobility are trapped by a single attractor. Our results reveal the passive fraction as a physically relevant control parameter in active systems, leading to rich self-organizing dynamics.
%
\end{abstract}

\maketitle

%
Active matter encompasses a broad class of systems composed of self-driven units that continuously consume energy to generate mechanical work, often expressed as directed motion~\cite{Vicsek2012, Romanczuk2012, Marchetti2013, Bechinger2016}.
These systems span a wide range of spatial and temporal scales, from natural settings including cytoskeletal assemblies, bacterial colonies, tissues, and animal groups such as bird flocks and herds, to artificial realizations such as vibrated granular rods, synthetic active colloids, or robotic systems~\cite{Bialek2012, Katz2011, Gomez-Nava2022, Buhl2006, Zhang2010, Beppu2017, Deseigne2010, Hamann2018, Ben2023}.
A defining feature of active matter is the emergence of collective phenomena far from equilibrium, including motility induced phase separation, flocking, and collective oscillations, which arise from an interplay between activity, interactions, and fluctuations~\cite{Cates2015, Vicsek1995, Toner1995, Ferrante2012, Ferrante2013NJP, Huepe2015, Khaluf2017, Henkes2020, Shea2025}.

Self-organization in active systems appears when disordered motion converges towards coherent collective dynamics, typically collective oscillations or flocking, with particles moving in parallel, coordinated trajectories.
It can emerge through different microscopic mechanisms, such as mutual-alignment forces between neighboring agents~\cite{Vicsek1995, Toner1995} or self-alignment forces resulting from agent-substrate interactions that orient each agent towards its displacement direction~\cite{Baconnier2025}.
While mutual-alignment has long been considered the right effective model for describing the convergence to flocking in dilute active systems, self-alignment has proven to be particularly relevant in dense and solid-like cases. Here, a combination of activity, elasticity, and self-alignment has been shown to lead to self-organization in a variety of biological and robotic contexts~\cite{Shimoyama1996, Szabo2006, Henkes2011, Ferrante2013, Zheng2020, Baconnier2022, Xu2023, Das2024, Baconnier2024, Baconnier2025, Arbel2025, Casiulis2025, Musacchio2026, Baconnier2024, Baconnier2025}.
%

Up to now, most research on active matter has focused on homogeneous systems of only active components. 
However, in many real-world realizations, active and passive constituents coexist, introducing heterogeneities that can strongly reshape the dynamics~\cite{Angelani2011, Buttinoni2013, Kummel2015, Wysocki2016, Smrek2017, Wittkowski2017, Vaccari2018, Sinaasappel2026}.
Although previous studies have analyzed the emerging phases in active--passive mixtures, including clustering, phase separation, and pattern formation~\cite{Schaller2010, Ibele2010, McCandlish2012, Palacci2013, Chepizhko2013, Mones2015, Dolai2018, Gokhale2022, Kushwaha2023, Jacob2025}, the effect of passive inclusions in the resulting collective states and their stability in dense systems is not well understood.
For mutual alignment, studies have shown that increasing the fraction of passive agents or fixed obstacles can destabilize the flocking order~\cite{Martinez2018, Bera2020},
but for self-aligning cases~\cite{Yllanes2017}, the effect of passive particle heterogeneity remains largely unknown.
%

Order-disorder transitions in active matter have been typically described as a function of parameters like noise strength, particle density, or activity level, which are hard to control in most real-world settings. 
An alternative and more experimentally viable approach is to control the fraction of passive constituents in the system, for instance by suppressing the activity of a subset of the agents.
This could occur naturally in biological systems as a result of cell death, dormancy, or metabolic arrest, and in technological settings due to inactive robots or other passive components.
Varying the passive fraction thus provides a direct, experimentally accessible way to control order-disorder transitions and emergent collective dynamics in heterogeneous active systems. 

\begin{figure}[!t]
\includegraphics[width=\linewidth]{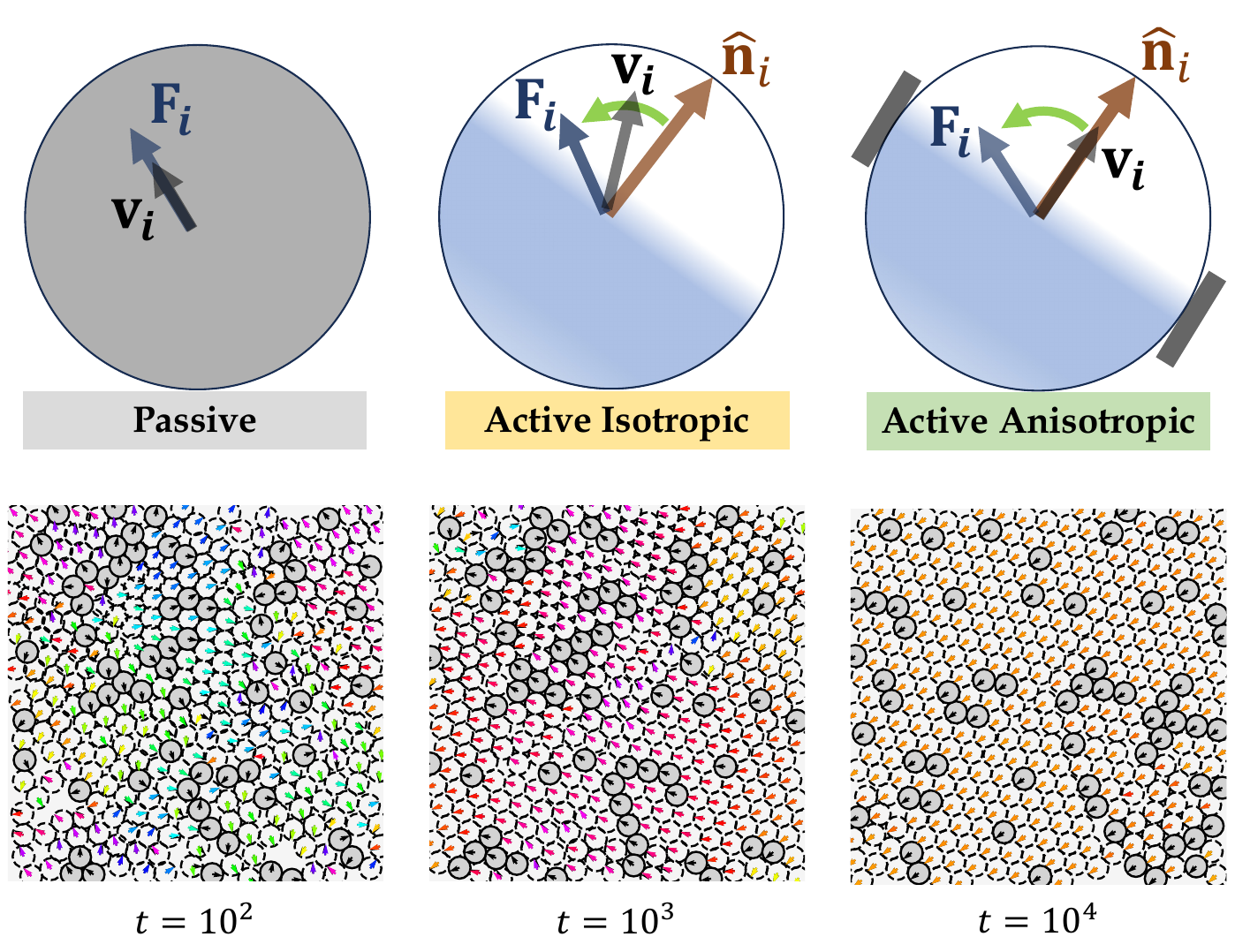}
\caption{Top: Schematic representation of the three types of disks in the studied models. Passive disks are pushed with velocity $\vv_i$ by external forces $\Fv_i$. Active isotropic disks move with $\vv_i$ resulting from a combination of self-propulsion along $\nv_i$ and external forces. 
Anisotropic disks can only move along $\nv_i$, as in wheeled vehicles. 
In both active cases, $\nv_i$ self-aligns towards $\vv_i$ (green arrows).
Bottom: Typical self-organizing dynamics in mixed systems of passive and (isotropic) self-aligning active particles with passive fraction $\alpha_p = 0.188$. Snapshots are displayed, at $t=10^2$, $10^3$, and $10^4$, showing the dynamics from local to global alignment.
Passive disks are shown in gray. Black arrows denote the velocity  directions $\hat{\vv}_i=\vv_i/|\vv_i|$; colored arrows show the $\nv_i$ directions.
}
\label{fig1}
\end{figure}

In this Letter, we investigate the emergent dynamics of dense mixtures of active self-aligning polar disks and passive disks.
Using the fraction of passive components as control parameter, we examine the corresponding order-disorder transition and the emergence of novel, partially-ordered, oscillating metastable states that depend on the spatial distribution of the passive disks and lattice defects.
We consider self-propelled disks with isotropic~\cite{Szabo2006, Das2024} or anisotropic~\cite{Zheng2020, Xu2023} disk-substrate damping interactions, showing that the properties of the transition and the metastable states change between these two types of damping.
Our findings shed light on the emergent states that could be experimentally observed in densely packed groups of active and passive components, with implications for understanding and controlling systems ranging from cellular assemblies to robotic swarms.

%
We begin by describing minimal models for the three types of disks considered: passive, active isotropic, and active anisotropic (see Fig.~\ref{fig1}).
The total force on any disk $i$ is defined by $\Fv_i  = \sum_{j}\fv_{ij}$, corresponding to the sum of its interactions with its $j$ neighbors through linear repulsive forces given by $\fv_{ij} = (k/l_0)(|\rv_{ij}|-l_0) \, \rv_{ij}/|\rv_{ij}|$
for $|\rv_{ij}| \leq l_0$, and by $\fv_{ij}=0$ otherwise.
Here, $k/l_0$ determines the repulsion strength, $\rv_{ij}$ is a vector between the disks, and $l_0$ is the disk diameter.
In order to analyze the emerging states in detail, we consider here noiseless systems and then show in the End Matter and Supplemental Material~\cite{Supply2026} that our results remain relevant when noise is added.
We also assume overdamped dynamics, a regime that describes most cellular and robotic systems of interest, although inertial effects can become important in certain realizations.
Under these conditions, the idealized dynamics of any passive disk $i$ will simply obey $\dot{\rv}_i=\mu \Fv_i$. 
On the other hand, the dynamics of the active agents also depend on their interactions with the substrate, which can produce disk-substrate damping with differing degrees of isotropy or anisotropy, leading to isotropic or anisotropic mobility.
In the fully isotropic case, the repulsive forces between agents can push them equally in any direction, as in the passive particle case, regardless of agent orientation. Instead, in the fully anisotropic case, they can produce no lateral displacement (no motion perpendicular to the heading), as in the case of wheeled vehicles (see Fig.~\ref{fig1}, top panels).
We consider both limit cases below, implementing simulations with positional dynamics given by one of the following two expressions~\cite{Baconnier2025}
\bea
\dot{\rv}_i &=&  v_0 \nv_i + \mu \Fv_i ~~~~~~~~~~~~~~~~({\rm Isotropic}) ~,\label{eom1}\\
\dot{\rv}_i &=&  v_0 \nv_i + \mupara \nv_i \nv^{T}_i \Fv_i ~~~~~({\rm Anisotropic})~.
\label{eom2}
\eea
Here, $v_0$ is the self-propulsion speed, $\nv_i$ is a unit vector pointing in the agent's heading direction, $\mu$ is the isotropic mobility coefficient, and $\mupara$ is the mobility coefficient along $\nv_i$, since the term with $\nv_i \nv_{i}^{T} \Fv_i = (\nv_i \cdot \Fv_i) \nv_i$ projects the total force in this direction~\cite{Ferrante2013, Tang2025}.
In addition, the angular dynamics is given by
\bea
\dot{\nv}_i &=& \b \left(\mathbb{I}-\nv_i \nv^{T}_i\right)\Fv_i\,,
\label{eom3}
\eea
where $\mathbb{I}$ is the $2\times2$ identity matrix and $\beta$ is the inverse rotational damping coefficient, which controls the degree of self-alignment.

We investigate mixtures of $N$ interacting particles composed of $N_a$ active and $N_p$ passive components ($N_a + N_p = N$). The relative abundance of passive particles is quantified by the passive fraction $\alpha_p = N_p/N$, which serves as the primary control parameter across all reported simulations.
The dynamics of the $N_a$ self-propelled polar disks are evolved using an Euler time-stepping scheme with time step $dt = 0.01$, integrating Eq.~\eqref{eom1} (isotropic case) or Eq.~\eqref{eom2} (anisotropic case), together with the orientation dynamics in Eq.~\eqref{eom3}. 

Each simulation is initialized by randomly placing the active and passive disks within the domain, assigning uniformly random $\nv_i$ orientations to the active agents. To ensure that our measurements reflect steady-state behavior, we discard the first $2\times10^7$ time steps of each run, then record $2\times10^3$ configurations sampled every $10^4$ time steps over an additional $2\times10^7$ time steps. 
Unless otherwise stated, all simulations are performed at fixed packing fraction $\Phi = 0.907$ within a periodic square domain of area $\pi N l_0^2 /(4\Phi)$. We use $N = 336$, mobility coefficient $\mu = 0.02$ (isotropic) or $\mu_{\parallel} = 0.02$ (anisotropic), repulsion strength $k = 5$, particle diameter $l_0 = 1.09$, self-alignment strength $\beta = 1.2$, and self-propulsion speed $v_0 = 0.002$. An example of the resulting dynamics is presented in Fig.~\ref{fig1} (bottom) for isotropic active disks with $\alpha_p = 0.188$, showing the emergence of global alignment starting from an initial disordered state.

\begin{figure}[!t]
\includegraphics[width=\linewidth]{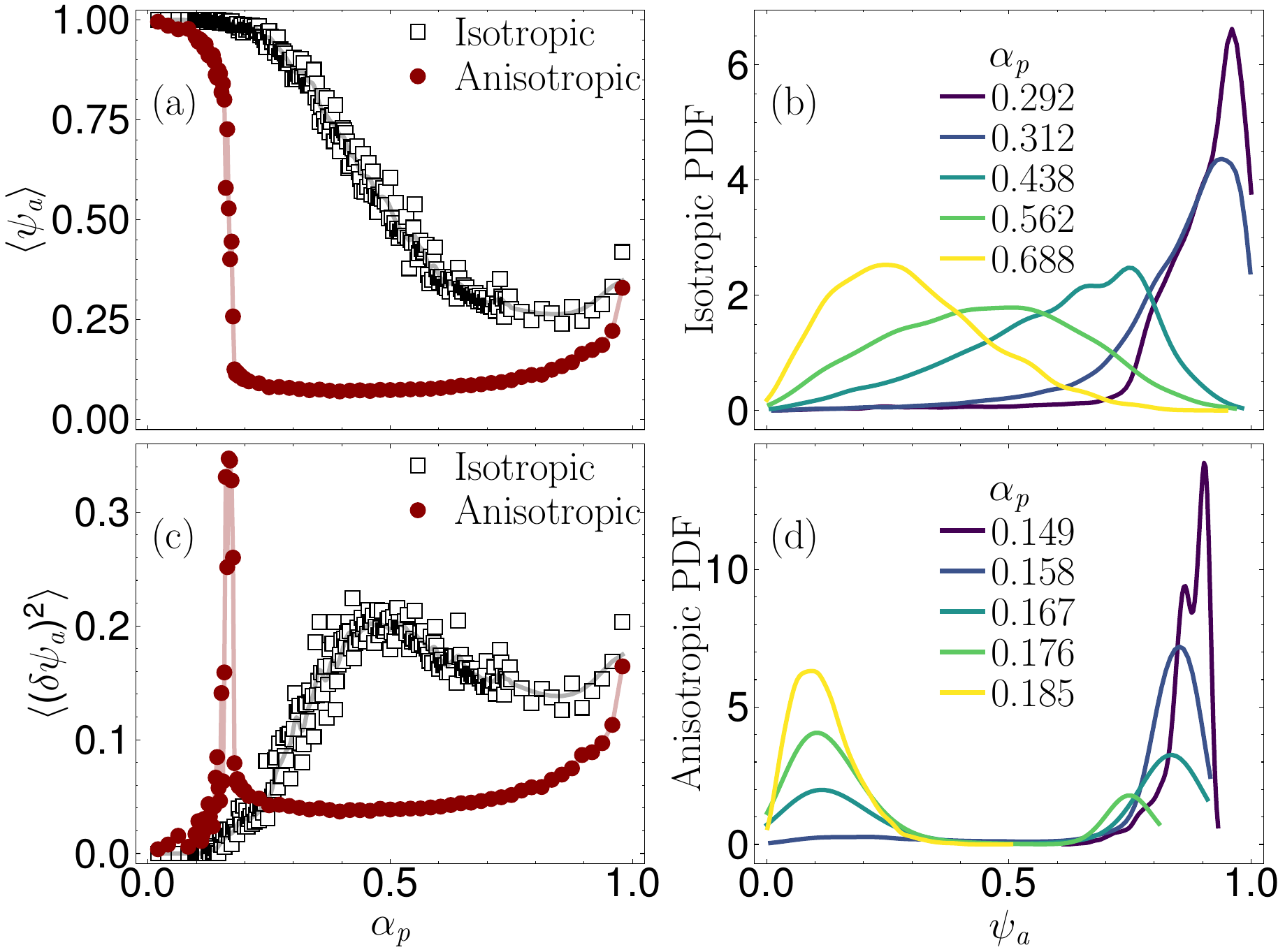}
\caption{Polarization statistics of active disks in active–passive mixtures as a function of passive fraction $\alpha_p$.
Left: Mean polarization $\langle \psi_a \rangle$ (a) and variance $\la(\delta \psi_a )^2\ra$ (c) for the isotropic (black squares) and anisotropic (red dots) mobility cases. 
Each symbol shows the average over 10 realizations, connected by lines in the anisotropic case and with an added sliding average curve in the isotropic one. 
Order is lost in both cases as $\alpha_p$ is increased, through a smooth transition for isotropic mobility and a discontinuous jump at lower $\alpha_p$ for anisotropic.
Right: Mean probability density functions for the isotropic (b) and anisotropic (d) mobility cases.
The isotropic case remains unimodal for all $\alpha_p$, with a peak slowly shifting from high to low $\psi_a$ as the passive fraction is increased.
The anisotropic case displays a bimodal state at intermediate $\alpha_p$, indicative of a discontinuous transition.
}
\label{fig2}
\end{figure}     

To evaluate the degree of order in the system, we compute the velocity polarization separately for the active and passive components as 
$\psi_{a,p} = \left|\sum_{i=1}^{N_{a,p}} \hat{\mathbf v}_i\right|/N_{a,p}$.
Here, $\psi_{a}$ and $N_{a}$ are the polarization and number of active particles, $\psi_p$ and $N_p$ are the corresponding passive ones, and $\hat{\mathbf v}_i = \mathbf v_i / |\mathbf v_i|$ is the velocity unit vector.
For each set of parameters, we obtain the mean polarization averaged over time,
$\langle \psi_{a,p}\rangle$, and corresponding mean variance 
$\la (\delta \psi_{a,p})^2\ra = \langle \psi_{a,p}^2\rangle-\langle \psi_{a,p}\rangle^2$.

Figure~\ref{fig2} presents the velocity polarization 
$\psi_{a}$ of the active disks as a function of the passive fraction $\alpha_p$, for isotropic and anisotropic mobility.
The equivalent statistics for passive disks exhibit qualitatively similar behavior, albeit less marked (see Supplemental Material Fig.~S.1~\cite{Supply2026}).
Panel (a) shows that increasing the passive fraction $\alpha_p$ drives the mean polarization to transition to a disordered state, through a continuous transition in the isotropic case and a discontinuous transition in the anisotropic case.
This is verified by the variance displayed in panel (c) and by the probability density functions (PDFs) in panels (b,d), which show unimodal distributions in the isotropic case and bimodal in the anisotropic one.
Simulations of larger systems show that both transitions sharpen as $N$ grows, consistent with genuine phase transitions that persist in the thermodynamic limit (see Supplemental Material Fig.~S.2~\cite{Supply2026}). In contrast, the increase in all curves that can be seen in panels (a,d) for $\alpha_p \rightarrow 1$ diminishes with increasing system size, indicating that this is a finite‑size effect produced by the vanishing number of active disks in this limit.

Interestingly, the change in the nature of the transition when switching from isotropic to anisotropic mobility can be captured at the mean-field level. Indeed, in this approximation the polarization obeys a Landau amplitude equation $\dot P = a_0 P - b P^3 + c P^5$, where increasing the passive fraction reduces the effective alignment gain and enhances crowding, here $b$ grows with $\alpha_p$.
This yields a continuous loss of order in the isotropic case, but a discontinuous one when anisotropy increases nonlinear amplification
(see End Matter for details). 

Although the analysis above assumes that an ordered state that includes passive components can be described as standard flocking, closer examination reveals additional structures in the polarization statistics that cannot be captured by a coarse‑grained description. 
Indeed, in both the isotropic and the anisotropic cases, the $\psi_a$ distributions in Figs.~\ref{fig2}(b) and (d) exhibit nontrivial secondary features: rather than a  simple single maximum, they often display weaker secondary peaks or oscillatory modulations, around the global maximum and in the distribution tails. These signatures appear because the systems do not relax to unique fixed-point states; instead, they can transition between metastable configurations and become trapped in attractors with strong oscillations.

In what follows, we will focus on understanding the origin of these secondary features. To do this, we will first examine the temporal dynamics of representative realizations in the isotropic case, and then extend our analyses to the anisotropic case and to  systems with noise (see End Matter Fig.~\ref{fig5}).
%

\begin{figure}[!t]
\includegraphics[width=\linewidth]{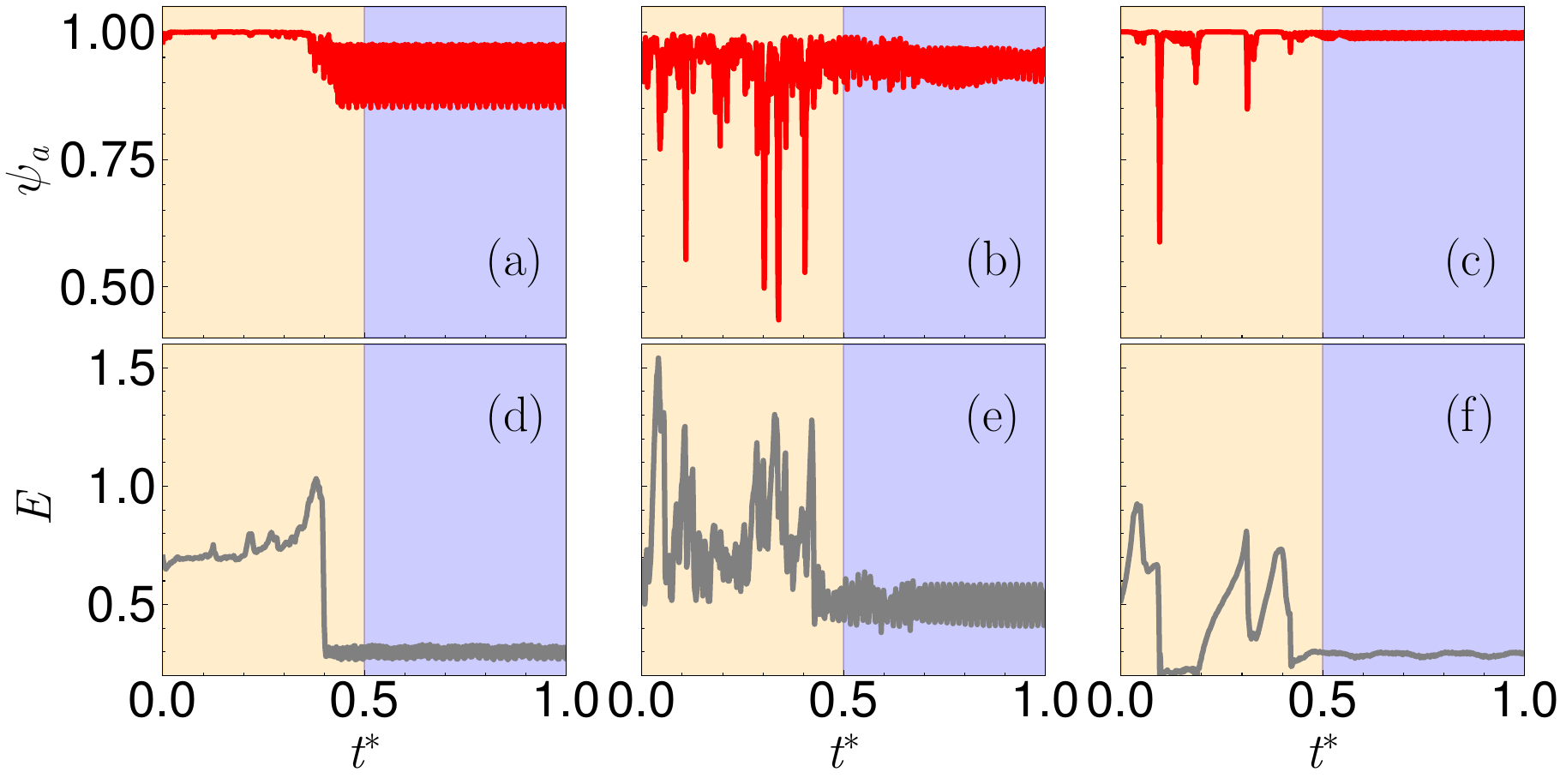}
\caption{Transition examples between metastable states in three simulation realizations of dense active–passive mixtures of isotropic mobility disks with passive fraction $\alpha_p = 0.188$.
Top panels display the polarization $\psi_a$, and bottom panels the elastic energy $E$ as a function of renormalized time 
$t^* = \left( t - t_0 \right) / 10^5$, with
$t_0=10^4$ (a,d),
$1.9\times 10^5$ (b,e), or
$1.7\times 10^5$ (c,f).
Elastic energy accumulates as $t^*$ approaches $0.5$, until the system transitions to a new metastable state with lower $E$. 
%
Panels (a,d) show a transition from a highly polarized state with disordered fluctuations to an oscillatory state with lower mean polarization.
Panels (b,e), from a disordered state to a rotating state with higher polarization.
Panels (c,f), from a highly aligned state with intermittent disorder to well-aligned ordered oscillatory dynamics.
}
\label{fig3}
\end{figure}    

Figure~\ref{fig3} shows three representative realizations that contribute to the polarization distributions displayed in Fig.~\ref{fig2}(b), corresponding to simulations in the isotropic mobility case and in the ordered regime ($\alpha_p=0.188$).
These examples show that transient and long-term dynamics vary markedly across realizations.
Panel (a) presents an initially ordered configuration that transitions to a long‑lived state with large-amplitude collective oscillations.
Panel (b) shows instead strongly irregular early fluctuations that eventually settle into a locally rotating, polarized state.
Panel (c) displays transiently ordered intermittent states that transition to  a highly polarized state with small collective oscillations.
Together, these behaviors demonstrate that isotropic mixtures near the transition support multiple competing dynamical attractors.
Despite these differences, the total elastic energy in panels (d-f) follows  similar qualitative dynamics across all realizations: the system accumulates mechanical tension due to self-propulsion and disk repulsion, which is then released through disk rearrangements that lower the elastic energy.

\begin{figure}[t]
\includegraphics[width=\linewidth]{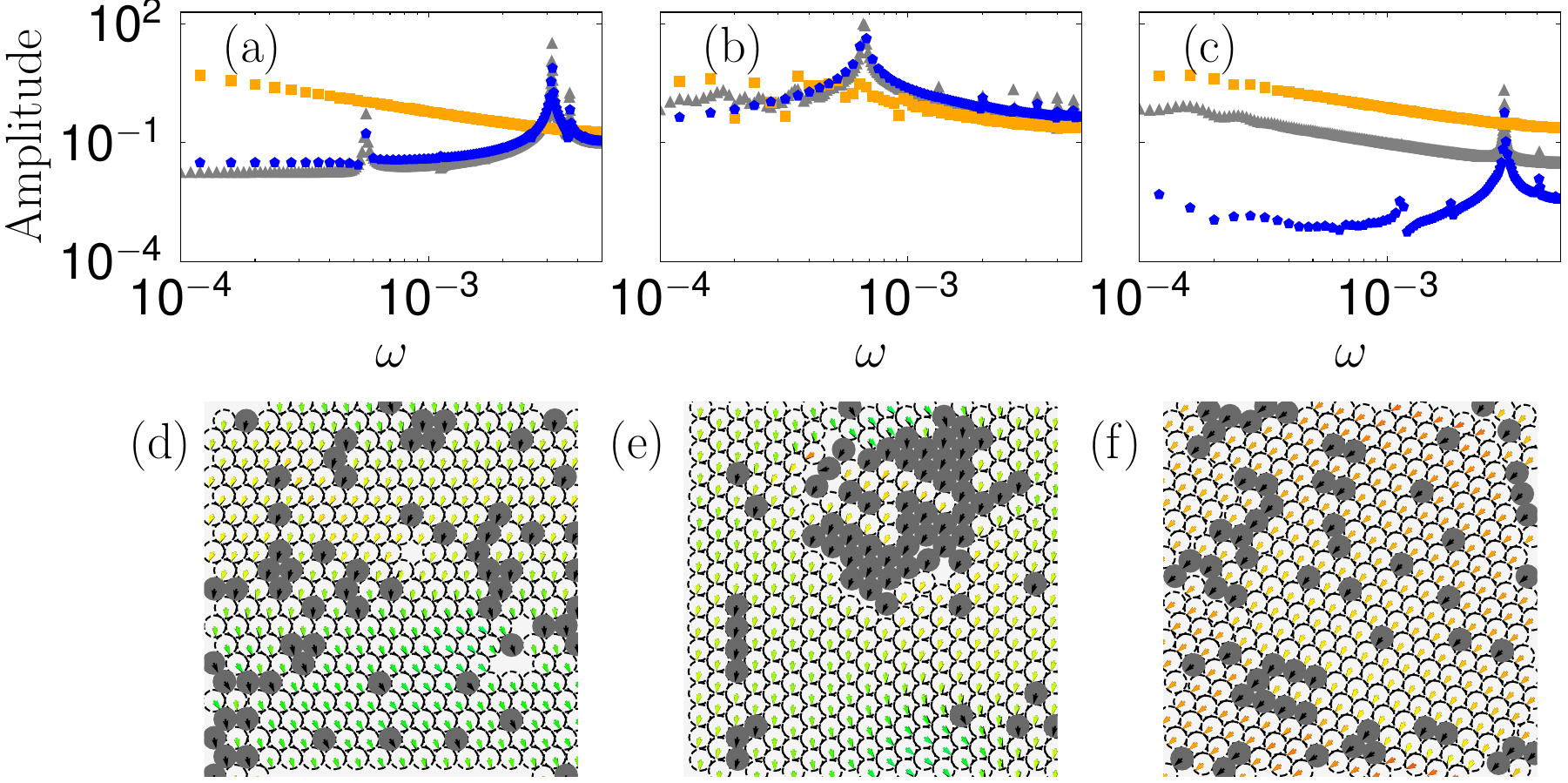}
\caption{Characterization of metastable states for the same simulations presented in Fig.~\ref{fig3}.
The top panels display the Fourier spectra of the normalized velocity autocorrelation, computed in the period 
before $0 \leq t^* \leq 0.5$ (orange squares) or 
after $0.5 < t^* \leq 1$ (blue pentagons) the transition, with the gray triangles covering the whole $0 \leq t^* \leq 1$ period.
The peaks in the spectra reveal the presence of oscillatory dynamics or localized rotation in the final sates reached by all realizations. 
The bottom panels present simulation snapshots of the final metastable states, using arrows to display the instantaneous disk velocity and showing passive disks in dark gray.
Localized rotation is reached in panels (b,e), instead of the collective translation shown in (a,d) and (c,f), due to the final distribution of passive particles, which cluster together into an effective barrier that hinders collective displacements.
}
\label{fig4}
\end{figure}

Figure~\ref{fig4} characterizes different metastable states that were visited in the realizations presented in Fig.~\ref{fig3}.
Panels (a-c) display the Fourier spectrum of the normalized velocity autocorrelation function 
$C(\tau)=\langle \hat{\mathbf v}_i(t)\cdot\hat{\mathbf v}_i(t+\tau)\rangle_{i,t}$, 
where $\hat{\mathbf v}_i=\mathbf v_i/|\mathbf v_i|$ and the mean is taken over all particles and over the selected time window. 
In these plots, well-defined spectral peaks indicate persistent periodic or quasi-periodic dynamics.
All three realizations exhibit metastable states with clear peaks showing different oscillation frequencies.
Panels (d-f) display the spatial distribution of passive particles corresponding to each of the three realizations.

Taken together, the realizations in Figs.~\ref{fig3} and ~\ref{fig4} (also included as Supplemental Material Movies~\cite{Supply2026}) show that the secondary features in the polarization distribution stem from the various metastable oscillatory states that emerge in active--passive mixtures.
The switching between these states can produce different polarization maxima, broadening the distribution, or a single periodic attractor with large oscillations that yields in itself a multimodal distribution.
We thus find that the self-organized regime in active–passive mixtures is given by a collection of periodic attractors near the aligned state, rather than by the single fixed point that defines the flocking state in the homogeneous case.

To exemplify the possible self-organizing dynamics, we now describe the metastable states in the selected realizations.
In the first run---panels (a,d), Movie 1---the system transitions from an aligned state with disordered oscillations to one with strong collective oscillations, undulatory trajectories, and lower elastic energy.
In the second run---panels (b,e), Movie 2---the system converges to a coherently rotating, aligned state, after starting in a disordered state that repeatedly approaches alignment but fails to sustain polarization.
In the third run---panels (c,f), Movie 3---an ordered state with irregular fluctuations and intermittent disorder transitions to an aligned state with small regular oscillations. 
We note that the metastable states that emerge will depend on the final distribution of passive components.
Indeed, in the second realization translational motion is suppressed in favor of localized rotation due to the agglomeration of passive particles into extended clusters that act as containment walls. Instead, in the first and third realizations collective translation is facilitated by the passive particles forming lanes along the direction of motion, which minimizes mechanical frustration.

If we now consider cases where the active disks have anisotropic agent-substrate interactions, we find that their mixtures also develop metastable oscillatory states in the ordered regime. As in the isotropic case, these are characterized by persistent oscillations in the polarization and by peaks in the normalized velocity autocorrelations (see Supplemental Material Fig.~S.5~\cite{Supply2026}).
However, in contrast to the isotropic case, switching between different ordered attractors within a single realization appears to be strongly suppressed. The mobility constraints imposed by fully anisotropic agent-substrate interactions, which prohibit lateral displacements, inhibit collective rearrangements and confine the dynamics to a single attractor over the duration of our simulations.

In sum, we have demonstrated that the passive fraction $\alpha_p$ in a dense mixture of self-propelled self-aligning disks and passive components serves as a  control parameter for its order–-disorder transition.
At fixed activity and packing fraction, two qualitatively distinct scenarios emerge depending on the active disk mobility: fully anisotropic mobility produces a discontinuous collapse of collective polarization; isotropic mobility yields a continuous transition and switching ordered states.
A mean-field Landau amplitude equation, derived from the microscopic heading dynamics, captures this dichotomy through the competing effects of alignment gains and passivity-induced crowding  (see End Matter for details).

In the ordered regime the system can self-organize into a  variety of long-lived metastable attractors that are not captured by coarse-grained descriptions.
These include aligned states displaying collective heading oscillations or localized rotation and are selected by the spatial distribution of passive particles and lattice defects.
In the isotropic case, different metastable states can be visited during each simulation. In the anisotropic case, lateral mobility constraints inhibit collective rearrangements, strongly suppressing the transition between attractors and confining the dynamics to a single attractor for the whole simulation.

We conclude by pointing out that, despite focusing here on the zero noise limit, our results remain valid for low levels of noise (see End Matter and Supplemental Material Fig.~S.4~\cite{Supply2026}). We thus expect them to be observable under real world conditions, in a variety of active biological or robotic systems with passive components.
Above a critical noise level, however, collective order will be directly destroyed by fluctuations~\cite{Szabo2006, Ferrante2013} through a standard noise-driven mechanism that is distinct from the passivity-driven transition studied here.

\medskip

\noindent
\textbf{Acknowledgments-}
W.T. and Z.H. acknowledge support from the National Natural Science Foundation of China (Grant No.\ 62176022).
Y.Z. and P.R. acknowledge support from the Deutsche Forschungsgemeinschaft (DFG, German Research Foundation) under Germany's Excellence Strategy (EXC 2002/1 ``Science of Intelligence,'' SCIoI), Project No.\ 390523135.
C.H. and A.S. acknowledge support from the John Templeton Foundation (Grant No.\ 62213).

\medskip

\noindent
\textbf{Data availability-}
The data that support the findings of
this article are openly available~\cite{Data2025}.

\medskip

\noindent
\textbf{Code availability-}
Simulation and analysis codes are available from the corresponding author upon reasonable request.

\bibliography{reference}

\section{End Matter}

\subsection{Mean-field transition}
We present a simple phenomenological mean-field description of self-alignment in dense mixtures that can help explain the change from continuous to discontinuous transitions observed when switching from isotropic to anisotropic disk mobility (see Fig.~\ref{fig2}).

Starting from the microscopic heading dynamics in Eq.~\eqref{eom3}, we write a kinetic equation for the one-particle heading angle distribution of the active disks and expand about the isotropic (disordered) state.
Projecting onto the first two angular harmonics then yields coupled evolution equations for the polarization vector $\mathbf{P}$ and the nematic tensor $\mathbf{Q}$ (see Supplemental Material~\cite{Supply2026}) given by~\cite{Bertin2006,Marchetti2013} 
\begin{align}
\dot{\mathbf{P}} &= \beta\left(\tfrac{1}{2}\mathbb{I}-\mathbf{Q}\right)\cdot\mathbf{F},
\label{eq:MT_p}\\
\dot{\mathbf{Q}} &= \frac{\beta}{2}\left[\mathrm{Sym}(\mathbf{F}\mathbf{P}^T)-\tfrac{1}{2}(\mathbf{F}\cdot\mathbf{P})\mathbb{I}\right]\,,
\label{eq:MT_Q}
\end{align}
where $\mathrm{Sym}(\mathbf{F}\mathbf{P}^T)=\tfrac{1}{2}(\mathbf{F}\mathbf{P}^T+\mathbf{P}\mathbf{F}^T)$
and $\mathbb{I}$ is the $2\times 2$ identity matrix.
At the mean-field level, dense repulsive contacts generate a net force that is collinear with the polarization, so we have $\mathbf F = F_0\,\mathbf P$.
Here the scalar prefactor $F_0$ encodes the competition between (i) \emph{alignment gain} produced by coherent active collision forces and (ii) \emph{passive inclusions} that oppose this coherent motion. This prefactor can be expressed in its general form for the cases with isotropic and anisotropic mobility as
\begin{equation}
F_0=
\begin{cases}
\dfrac{v_0}{\mu} \left(C_a \alpha_a-C_p\alpha_p\right), & \text{isotropic},\\[6pt]
\dfrac{v_0}{\mupara} \left(\chi_\parallel C_a^\parallel \alpha_a - C_p^\parallel \alpha_p\right), & \text{anisotropic}\,.
\end{cases}
\label{eq:F0}
\end{equation}
Here $\alpha_a=N_a/N$ and $\alpha_p=N_p/N$ are the corresponding active and passive fractions.
The coefficients $C_a$ and $C_p$ (and their longitudinal counterparts in the anisotropic case) represent the average contributions of active and passive neighbors to the mean contact force in a dense packing.
In the anisotropic model, we introduce an additional prefactor $\chi_\parallel$ because translation is restricted to the heading direction, so only the longitudinal component of the contact force contributes.
When averaged over all $\theta$ directions, this gives $\chi_\parallel=\langle \cos^2\theta\rangle=1/2$, which reduces the effective alignment gain relative to isotropic mobility.

Linearizing Eqs.~\eqref{eq:MT_p} and \eqref{eq:MT_Q} 
about $\mathbf{P}=\mathbf 0$ (with $\mathbf{Q}=\mathbf{0}$) gives 
$\dot{\mathbf{P}}=(\beta F_0/2)\mathbf{P}$.
Thus, loss of order with increasing 
$\alpha_p$ corresponds to a sign change of the effective drive $F_0$.
Setting $F_0=0$ in Eq.~\eqref{eq:F0} to find the critical passive fraction 
$\alpha_p^*$ gives $\alpha_p^* = C_a\alpha_a/C_p$ (isotropic) and 
$\alpha_p^* = \chi_\parallel C_a^\parallel\alpha_a/C_p^\parallel$ (anisotropic).
Since $\chi_\parallel = 1/2$, the anisotropic transition occurs at roughly 
half the passive fraction of the isotropic one (with crude assumption of comparable force 
coefficients $C_a\approx C_a^\parallel$, $C_p\approx C_p^\parallel$), 
consistent with the approximately twofold separation of the two transition 
points visible in Fig.~\ref{fig2}(a,c).

To capture the saturation and eventual collapse observed in simulations, we incorporate the leading feedback of dense crowding by allowing the effective force scale to decrease with increasing order, $F_0(P)=\tilde F_0\,[1-sP^2+tP^4+\cdots]$ with $s=\eta\alpha_p>0$.
Physically, $\eta$ quantifies passive-induced crowding/jamming that penalizes coherent motion at large polarization.
Substituting $\mathbf{F} = F_0(P)\mathbf{P}$ into Eq.~\eqref{eq:MT_p} yields a Landau amplitude equation~\cite{Cross1993} given by
\begin{equation}
\dot{P} = a_0 P - b\,P^3 + c\,P^5 + \mathcal{O}(P^7),
\label{eq:Landau}
\end{equation}
with $a_0=\beta\tilde F_0/2$, $b=(\beta/2)\tilde F_0\,s$, and $c=(\beta/2)\tilde F_0\,t$.
Increasing $\alpha_p$ therefore acts in two coupled ways: it reduces the linear alignment gain through $\tilde F_0\propto(C_a\alpha_a-C_p\alpha_p)$, while simultaneously strengthening the dominant nonlinear crowding correction through $s=\eta\alpha_p$.
In isotropic mobility, $a_0$ and $b$ decrease together as $\alpha_p$ 
increases: $a_0 \propto \tilde{F}_0$ shrinks because passive inclusions 
reduce the net alignment gain, while $b = (\beta/2)\tilde{F}_0\,\eta\alpha_p$ 
first grows (rising $\alpha_p$) and then falls (falling $\tilde{F}_0$), keeping 
$b^2 < 4a_0 c$ throughout.
The ordered state $P^* = \sqrt{a_0/b}$ therefore 
decreases smoothly to zero only when $a_0$ changes sign, producing the continuous suppression of polarization seen in Fig.~\ref{fig2}(a).
In anisotropic mobility, the reduced alignment gain (via $\chi_\parallel = 
1/2$) lowers $a_0$ more rapidly with $\alpha_p$, while the stronger effective 
crowding amplification (larger $\eta$) increases $b$ more quickly, so that 
$b^2 > 4a_0 c$ with $t > 0$ is satisfied before $a_0$ changes sign.
Under this condition the potential 
$V(P) = -(a_0/2) P^2 + (b/4) P^4 - (c/6) P^6$ 
develops a local minimum at $P>0$ that 
disappears through a saddle-node bifurcation, so the ordered branch loses 
stability discontinuously, producing the abrupt collapse of polarization 
observed in Fig.~\ref{fig2}(a).

\begin{figure}[!t]
\includegraphics[width=\linewidth]{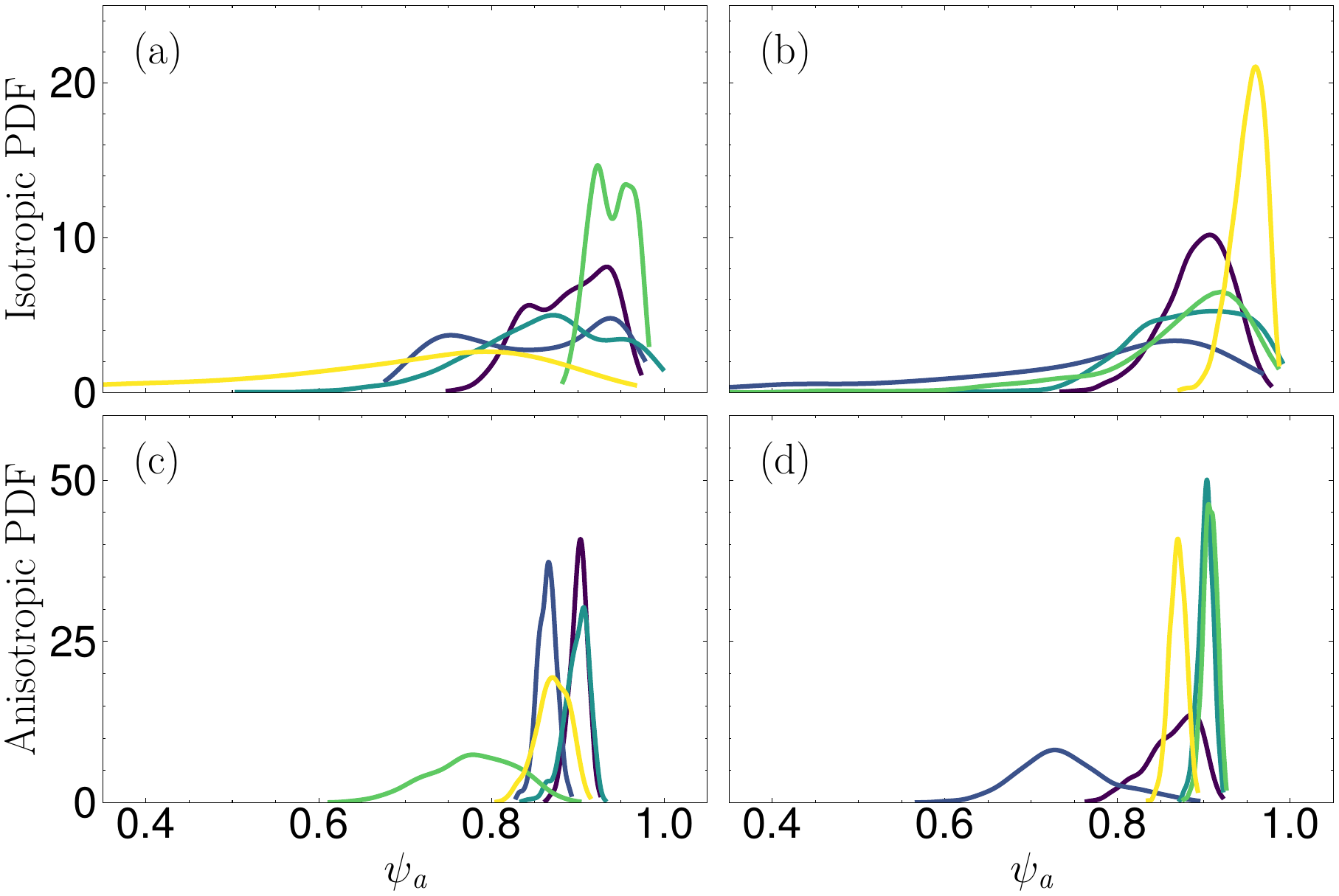}
\caption{Probability density functions of the 
polarization of active disks $\psi_a$ for five independent realizations in cases with isotropic or anisotropic mobility (top-bottom rows) and in the noiseless limit or with finite rotational noise (left-right columns).
Noise is set to $D_\theta=0.001$ in cases where it is present.
The passive fractions $\alpha_p$ are chosen to place the system in the ordered phase but close to the transition point, so that the mean polarization $\langle\psi_a\rangle$ is high while metastable fluctuations are pronounced.
We thus set 
$\alpha_p=0.312$, $\alpha_p=0.250$, $\alpha_p=0.149$, $\alpha_p=0.104$ 
in panels (a), (b), (c), (d), which correspond to 
$\langle\psi_a\rangle = 0.870$, $\langle\psi_a\rangle=0.897$, $\langle\psi_a\rangle=0.865$, $\langle\psi_a\rangle=0.877$,
respectively.
}
\label{fig5}
\end{figure}

\subsection{Impact of noise and anisotropy}
To further assess the roles of noise and mobility anisotropy in the emergent metastable dynamics, we examine the PDFs of the velocity polarization for individual realizations, after they reach a steady state in the ordered regime. Figure~\ref{fig5} shows the PDFs of five independent runs using active agents with isotropic or anisotropic mobility (top and bottom panels, respectively), in cases with and without noise (left and right panels).

In the noiseless case, systems with isotropic mobility [Fig.~\ref{fig5}(a)] exhibit broad distributions that vary markedly from realization to realization and typically display multiple maxima.
By contrast, the fully anisotropic case [Fig.~\ref{fig5}(c)] also shows realization-dependent distributions, but these are generally narrower and characterized by a single dominant peak.
This qualitative difference can be traced to the stronger mechanical constraints imposed by anisotropic mobility, due to the suppression of lateral displacements. 
This forces anisotropic realizations to rapidly lock into the first ordered attractor they encounter, staying confined to it throughout the simulation rather than exploring different metastable states that contribute to broader distributions with additional features.
Moreover, the restricted mobility suppresses ordered states with large-amplitude oscillations, which also limits the width of the resulting distributions.

We next examine the effect of noise by adding to Eq.~\eqref{eom3} a stochastic rotational term, $\sqrt{2D_{\theta}} \xi_i(t) \hat{\mathbf n}_{i}^{\perp}$, where $D_{\theta}$ is the angular diffusion constant and $\xi_i(t)$ is a random variable that introduces Gaussian white noise with zero mean and delta correlations $\langle\xi_i(t)\xi_j(t')\rangle=\delta_{ij}\delta(t-t')$~\cite{Szabo2006,Ferrante2013}.
In the presence of noise, both the isotropic and the anisotropic cases will converge to, and remain trapped in, a variety of long-lived metastable oscillatory states. The resulting polarization distributions exhibit a single dominant maximum for all noisy dynamics, in contrast to the noiseless isotropic case. This is because rotational noise suppresses prolonged trapping in distinct metastable states within a single realization, as well as attractors with large-amplitude oscillations, both contributors to the multimodal distributions observed in the absence of noise.
This behavior is consistent with the well-known role of noise in smoothing multistability and weakening trapping, here associated to the oscillatory and rotational attractors. Importantly, we find that the diversity of final metastable oscillatory states persists even at nonzero noise, indicating that these dynamical regimes should remain observable in experimental realizations, where noise is unavoidable.

\end{document}